# TRUNCATED HORSESHOES AND FORMAL LANGUAGES IN CHAOTIC SCATTERING

G. TROLL

Techn. Universität Berlin, FB Mathem. MA 7-2, D-10623 Berlin, Germany
*E-mail:* troll@math.tu-berlin.de, *FAX:* GER + 30 - 314 21577

ABSTRACT. In this paper we study parameter families of truncated horseshoes as models of multiscattering systems which show a transition to chaos without losing hyperbolicity, so that the topological features of the transition are completely describable by a parameterized family of symbolic dynamics. At a fixed parameter value the corresponding horseshoe represents the set of orbits trapped in the scattering region. The bifurcations are a pure boundary effect and no other bifurcations such as saddle center bifurcations occur in this transition scenario. Truncated horseshoes actually arise in concrete potential scattering under suitable conditions. It is shown that a simple scattering model introduced earlier can realize this scenario in a certain parameter range (the "truncated sawshoe") . For this purpose, we solve the inverse scattering problem of finding the central potential associated to the sawshoe model. Furthermore, we review classification schemes for the transition to chaos of truncated horseshoes originating from symbolic dynamics and formal language theory and apply them to the truncated double horseshoe and the truncated sawshoe.

This work was supported by the Deutsche Forschungsgemeinschaft within a project of the Sonderforschungsbereich 288.

1. INTRODUCTION

We are going to study the following type of scattering systems. A set of scatterers is arranged in some region of the real plane according to some deterministic law. A test particle is injected into this scattering region and will undergo multiple scatterings before (if ever) it will escape. A discretization of this dynamics is assumed to be possible in the form of a map $T$, which describes a single scattering event. Certain features of this scattering system such as the singularities of reaction functions or cross sections, can be described by the set of trapped trajectories which never escape from the scattering region. Introducing and varying an external parameter, controlling the potential strengths say, varies the trapped set of the scattering system. In 2 D the trapped set typically takes the form of a Smale's horseshoe.

In this paper we suggest the following abstraction. We interpret parameter families of truncated horseshoes as models of multiscattering systems which show a transition to chaos completely describable by a parameterized family of symbolic dynamics[1] (SD). At a fixed parameter value the corresponding horseshoe represents the trapped set. Deforming or chopping the horseshoe will change the trapped set and consequently the physical properties of the whole system.

A horseshoe can be generated or destroyed in two ways: either by chopping off parts of it or by smoothly deforming and disentangling the horseshoe tangle yielding *truncated horseshoes* and *disentangled horseshoes*, respectively. (cf. figure 1). The property distinguishing these two scenarios is that truncation of a horseshoe preserves the hyperbolicity[2] of the underlying dynamical map, whereas smooth deformations produce tangent bifurcations with the consequent loss of hyperbolicity. This is why truncated horseshoes are the simpler models, which lend themselves more readily to a rigorous approach.

Disentangled horseshoes as models for the transition to chaos of *potential scattering* have been discussed by Bleher, Ding, Grebogi, Ott, and Yorke in a series of papers (cf. [2, 1, 6]). But truncated horseshoes also arise in concrete potential scattering, if one chooses a suitable arrangement of non-overlapping, attractive potentials with compact supports (i.e. of finite range) and permits non-differentiability of the potentials at the boundary of their supports. A sim-ple example of this kind (cf. [15]–[17] ) consists of such potentials which are additionally identical and central with a (single) sawtooth as deflection function and which are arranged at unit distances along a straight line. The potentials are parametrized by the slope $k > 0$ of the deflection function and by the range parameter $0 < b_{\max} < 0.5$.

---

[1] Some basic concepts about SD can be found in appendix A

[2] A set $\Lambda$ is called hyperbolic for a map or semi-flow $T$ if it is compact, invariant under $T$ and its tangent bundle can be continuously decomposed into a direct sum of stable and unstable manifolds, whose contraction factors have a uniform bound $< 1$.

FIGURE 1. (a) The time 2 and -2 image of a complete Smale horseshoe together with a coding for the vertical (forward) and horizontal (backward) strips. (b) A smooth deformation yielding an incomplete Smale horseshoe, where not all forward strips intersect all backward strips. (c) A Smale horseshoe truncated by a family of level lines of a pruning function $f$. The shaded convex region is supposed to satisfy $f(u) > \kappa$. The same 4 second generation rectangles are lost as in Fig. 1 (b). The forbidden symbol combinations are $\{0001, 1001, 1011, 0011\}$.

We shall show in the present paper that a sawtooth deflection function indeed arises from a suitable attractive central potential. The jump discontinuity of the sawtooth proves to have no severer effects than different right and left derivatives of the potential at the boundary of its support. Everywhere else the potential is smooth. Moreover, the trajectories are everywhere once continuously differentiable.

Actually, this model also has a non-hyperbolic chaotic regime for $0 < k < 4$, where the horseshoe develops by disentangling (cf. [11]). What we are interested in here however, is the parameter range for which a transition to chaos occurs within the hyperbolic regime ($b_{\max}$ small enough and $k > 4$). One must be aware that hyperbolicity alone does not assure structural stability and thus does not exclude bifurcations. In the systems we are concerned with here, there are boundary effects originating from the non-differentiability of the attractive potentials on the boundary of their support. These boundary effects generate the bifurcations leading to a transition of the dynamics from regular to chaotic behaviour (see next section). Only for the purpose of analysing the transition more precisely, we shall introduce a linearization. The qualitative features of the linearized version and the original model are the same. As the horseshoe associated to this model resembles a truncated sawtooth in the asymptotic limit and in the linearized model, we speak of a "sawshoe" in order to distinguish this particular model.

The advantage of truncated horseshoes is that their transition to chaos reduces to the problem of analysing the evolution of the trapped set in terms of the evolution of an associated SD.

In the family of associated SD the evolution of the trapped set occurs as an evolution of the set of "permitted" symbol sequences each describing an orbit of the trapped set. This set of permitted symbol sequences is called a *subshift*. If it comprises all symbol sequences it is called a *full shift*. As one often represents a subshift as a tree of permitted sequence segments, the evolution of subshifts corresponds to an evolution of trees called *pruning process*. Multiscattering families which are described by a pruning process will be called *pruned systems*.

We have identified two scenarios by which pruned systems can become chaotic:

a transition via a cascade of bifurcations and an abrupt transition.

Both, chaoticity and topological entropy of a pruned system essentially measure its size (cf. definition 3.4 and remarks there). Not even topologically do they determine the pruned system completely. In symbolic dynamics, there is a further refinement of classification by the Markov property of permitted sequences, namely into finite and infinite subshifts.

A means to classify infinite subshifts further is provided by formal language theory. A formal language is just a set of finite strings (as opposed to sequences in the case of subshifts) of symbols taken from a finite set called the alphabet. By relating formal languages with symbolic dynamics one can use formal language theory to study the structural complexity of pruned systems. Structural complexity means roughly the type of rules needed to describe the production of those words belonging to the formal language in question. The theory of automata and formal languages shows that there are types of rules which are not arbitrary but have certain universality properties. The best known examples are the Turing machines and the finite state automata, both of which belong to the Chomsky hierarchy of automata (or equivalently of languages). Important now for the applicability to dynamical systems is that the Chomsky hierarchy offers a classification scheme, which proves to be a also a topological invariant, i.e. topologically equivalent SD fall in the same complexity class.

There is now a bridge between structural complexity and the onset of chaos in the following sense: The onset of chaos in pruned systems shows certain features of a phase transition, seen for instance in the topological entropy as a function of the family parameters (cf. [16]). Recently there have been a few studies about the complexity theoretic description of phase transitions based upon the informational diversity and computational complexity of observed data (cf. [4]). It was suggested there that it is at phase transitions that high level[3] computation occurs, so that computational ideas provide a new set of tools for investigating the physics of phase transitions.

**1.1. Overview.** The contents of this paper is as follows: Section 2 explains, why truncated horseshoes can model chaotic scattering. Two simple examples are treated, the double horseshoe and the sawshoe. This part is based on [15] and [17]. A continuous and but for the boundary of its support smooth potential associated to the sawshoe model is given. This is new. The details are worked out in appendix B.

---

[3]e.g. in the sense of the Chomsky hierarchy.

Sections 3 and 4 review parts from [17], [19] and [10] for section 4.1. Section 3 introduces formal multiscattering systems (FMS) (definition 3.1) and quotes sufficient conditions for them to have a symbolic dynamics (theorem 3.1). Truncation of FMS is then determined by a set of level lines of a cut-off function. Whenever a particle overjumps such a level line it is supposed to escape from the scattering region (definition 3.3). Furthermore, two scenarios for transitions to chaos of truncated horseshoes are discussed in this section: the cascade transition with infinitely many bifurcations preceding and accumulating at the critical point where chaos sets in (this transition bears a strong resemblance to the period-doubling scenario of bounded dynamical systems) and the abrupt transition where "$1 + \epsilon$" bifurcations suffice to generate chaos, i.e. the dynamics restricted to the trapped set is not yet sensitively dependent on initial conditions at the critical value, but at any $\epsilon$ beyond it is chaotic. These concepts are applied to the double horseshoe and the sawshoe, which are shown to follow the cascade and the abrupt transition, respectively (corollary 3 and theorem 3.2).

Section 4 discusses the same transitions from the point of view of complexity theory. To this purpose the notion of dynamically generated languages is introduced. These are languages which can be naturally associated to subshifts (proposition 4). The theory of formal languages offers the concepts to grasp the observation that typically the rules specifying the permitted sequences in a pruning process become more and more complicated close to the critical point of a dynamical system, indicating the need of a higher level organization of these rules. The sine qua non for the application of theses ideas from formal language theory is given in theorem 4.1. This theorem essentially states that the complexity measure most often used in formal language theory, namely the Chomsky hierarchy, has an intrinsic dynamical meaning, i.e. is not an artifact of the special description, which means here the chosen symbolic dynamics. The complexity transitions of the two example horseshoes are described by corollary 4 and theorem 4.3.

Appendix A and C contain some basic concepts from symbolic dynamics and automata theory, respectively; in appendix B a potential with sawtooth deflection function is calculated and some of its properties, such as uniqueness, are discussed. A Method from classical inverse scattering theory (cf. [12]) is used.

**1.2. Acknowledgement.** I would like to thank Uzy Smilansky and Andreas Knauf for discussions.

## 2. Truncated horseshoes as scattering models

**2.1. The truncated double horseshoe.** One of the first models studied in the days of the rediscovery of chaotic scattering by physicists in the eighties (cf. [7]), was the three disk model: A particle is scattered in 2 D at three hard disks arranged at the vertices of an equilateral triangle. When one injects a whole bunch of particles coming from an interval which together with a given direction of motion form a strip enveloping one of the disks, one notices that those particles staying in the scattering region of the three disks for another scattering incident originate from two smaller subintervals. These subintervals in turn contain two still smaller subintervals each, which represent particles realizing at least one more, that is at least three, scatterings. Iteration of this procedure generates the Cantor set of initial conditions, for which particles will be trapped forever in the three disk system.

In potential scattering the more natural space to study is phase space. We restrict our attention to 2 D, The canonical dynamics associated to a Cantor set in 2D is Smale's horseshoe. Now, in order to find the most simple multiscattering system, we are going to turn the problem upside down:
*Is there a (natural ?) way to interpret the abstract Smale horseshoe as a multiscattering dynamics ?*

I suggest the following interpretation: Take the unit square $Q_1$ as representing the part of phase space which has the right properties to be scattered at least once at a single potential of some hypothetical arrangement of potentials. We shall call this part of phase space here *fundamental domain*. Let the dynamics of this single scattering be given by Smale's horseshoe map $T_1^s$, which stretches and folds the unit square back onto itself. Those parts of the square's first image that are mapped outside the square (for example the bent part of the horseshoe), is supposed to represent the phase space points of particles which have escaped the scattering region for good. Whereas the two strips within the unit square represent particles which will achieve at least one more scattering. Iteration yields the forward trapped set of all points which have undergone an infinite number of scatterings. Backward images yield in the same way the backward trapped set of all points which will undergo infinitely many scatterings. The Cantor set of Smale's horseshoe is then just the intersection of the forward and backward trapped set, called *the trapped set* $\Lambda(T_1^s, Q_1)$. By labeling the two strips of the horseshoe, its Cantor set can be identified with the set of all sequences with 2 symbols, say 0 and 1, the so-called full shift of two symbols.[4]

The arguably simplest way to generate more complicated trapped sets than the full shift of two symbols, is by truncating the unit square horizontally at a point $0 \leq y_{\max} \leq 1$. The new fundamental domain is therefore:

$$\text{(1)} \qquad Q_{y_{\max}} = \{(x,y) \in [0,1]^2 ; 0 \leq y \leq y_{\max}\}$$

Points mapped outside $Q_{y_{\max}}$ are supposed to have escaped from the scattering region. The resulting map is denoted by $T^s_{y_{\max}}$. The lowering of $y_{\max}$ below 1 has the effect that not all sequences of two symbols correspond to trapped orbits : this is called pruning. The question which will occupy us throughout this paper, reads as applied to the truncated horseshoe: *How can the trapped set* $\Lambda(T^s_{y_{\max}}, Q_{y_{\max}})$ *be characterized as a function of the family parameter* $y_{\max}$.

The truncated Smale's horseshoe is not yet the most simple imaginable multiscattering system, because it contains a reflection.[5] We therefore take instead (see figure 2) Smale's double folded horseshoe map $T_{y_{\max}}$ (which as it happens was the original form found by Smale). The double horseshoe family with trapped set

$$\Lambda(T_{y_{\max}}, Q_{y_{\max}}) = \bigcap_{i=-\infty}^{\infty} T^i_{y_{\max}}(Q_{y_{\max}})$$

$$= \{u = (x,y) \in Q_1 : \pi_y(T^i_1(u)) \leq y_{\max} \; \forall \, i \in \mathbb{Z}\}$$

where $\pi_y$ is the projection onto the $y$-component, is equivalent to the symbolic dynamics $(\sigma, \Sigma_\nu)$ over the alphabet $\mathring{A} = \{0,1\}$ with trapped subshift $\Sigma_\nu$ and truncation parameter $\mu \in \mathring{A}^{\mathbb{N}_0}$, determined by the fundamental domain:

$$\text{(2)} \qquad \Gamma_\mu = \{a \in \mathring{A}^{\mathbb{Z}} : P(a) \leq \mu\}$$

and the truncation (pruning) function

$$\text{(3)} \qquad P(a) = \sum_{i=0}^{\infty} a_i \, 2^{-i}$$

For an orbit $\mathbf{a} \in \mathring{A}^{\mathbb{Z}}/\sigma$ we define

$$\text{(4)} \qquad P(\mathbf{a}) := \sup_{a \in \mathbf{a}} P(a)$$

---

[4]Some pertaining basic definitions of finite, infinite, cyclic subshifts and (semi–) conjugacies can be found in the appendix.

[5]The eigenvalues are negative in one strip and positive in the other.

FIGURE 2. (a) The twice folded, piecewise linear horseshoe map applied to the unit square $Q = Q_1$. The symbolic dynamics chosen uses the first generation horizontal strips. (b)The horizontally truncated double horseshoe. As the reflection has been removed, the 4 second generation rectangles lost for the chosen cut-off parameter $y_{\max}$ correspond to the symbol combinations $0011, 0111, 1011, 1111$, respectively.

Observe that we interpret sequences with nonnegative indices such as $\mu$ as binary numbers, i.e. in the base 2.

## 2.2. The truncated "sawshoe" as a potential scattering model with complete SD:. 

We address in this section a multiscattering model introduced by the author and U. Smilansky in [15]. Its salient feature is that it yields a parameterized family of discrete dynamical systems which remain expansive throughout a transition to chaos. Therefore this transition to chaos can be described at least topologically in a complete way by a family of symbolic dynamics (cf. theorem 3.1 of section 3).

Suppose a particle of a fixed asymptotic energy $E_0 > 0$ is scattered at an attractive central potential $V = V^{(1)} \leq 0$. For the sake of simplicity we assume $V$ to be of finite range $b_{\max} > 0$. Conservation of angular momentum restricts the dynamics to a plane in configuration space. The equations of motion can be integrated to yield the polar angle $\phi$ of the trajectory as a function of the impact parameter $b$ and the radial distance $r$. The asymptotic final angle $x_f^{(1)}$ (for $t \to \infty$) is expressible as a function of the asymptotic initial angle $x_i^{(1)}$ (for $t \to -\infty$) and the impact parameter as:

$$(5) \qquad x_f^{(1)} = x_i^{(1)} + \theta_V(b)$$

where the deflection function $\theta_V$ is given by equation (31). Multiscattering can then be achieved by positioning in configuration space copies $V^{(i)}$ of $V$ according to some group action and avoiding overlap. The particle ejected from $V^{(1)}$ may either escape the scattering region for good or it may approach the next potential $V^{(2)}$ under a new initial angle and a new impact parameter both of which are functions given by the geometry of the arrangement of the potentials. The map $(x_i^{(1)}, b^{(1)}) \mapsto (x_i^{(2)}, b^{(2)})$ is then the discretized scattering map.

Placing the potentials along a line at equidistant positions yields the following scattering map:

$$(6) \qquad \begin{aligned} T \,:\, S^1 \times \mathbb{R} &\longrightarrow S^1 \times \mathbb{R} \\ \begin{pmatrix} x \\ b \end{pmatrix} &\longmapsto \begin{pmatrix} x + \theta_V(b) \\ b + G(x + \theta_V(b)) \end{pmatrix} \end{aligned}$$

where $S^1$ is the unit circle and

$$(7) \qquad G(x) = -\operatorname{sgn}(\cos x) \sin x$$

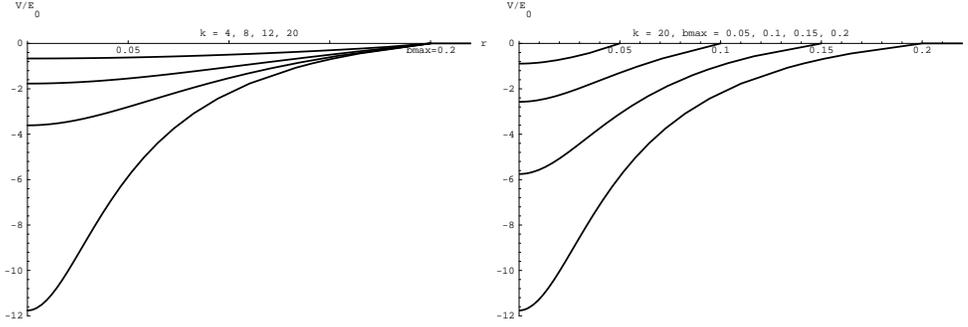

FIGURE 3. The potential $V_{Z_K}$, which has the sawtooth $Z_k$ as deflection function for some $b_{\max}$- and $k$-values.

In order to arrive at a concrete model we choose as deflection function $\theta_V$ the sawtooth function $Z_k$, $k > 0$:

$$(8) \qquad Z_k \colon \mathbb{R} \to \mathbb{R}, \quad b \mapsto \begin{cases} kb & \text{if } |b| \leq b_{\max} \\ 0 & \text{otherwise} \end{cases}$$

In the papers quoted the question whether there is indeed a potential with such a deflection function was left open. In the appendix this question is answered positively. It is shown that $Z_k$ is the deflection function of a potential $V_{Z_K}$ which is smooth in $]0, b_{\max}[$ and continuous on $\mathbb{R}_0^+$. The potential $V_{Z_K}$ is defined by the following parameter representation:

$$(9) \qquad \left\{ \left( s\, e^{-k/\pi\,\sqrt{b_{\max}^2 - s^2}},\, E_0\bigl(1 - e^{2k/\pi\,\sqrt{b_{\max}^2 - s^2}}\bigr) \right);\, s \in [0, b_{\max}] \right\}$$

together with $V_{Z_K}(r) := 0$ if $r > b_{\max}$. Figure 3 shows this potential for some values of $k$ and $b_{\max}$. Some properties are discussed in appendix B. Figures 4

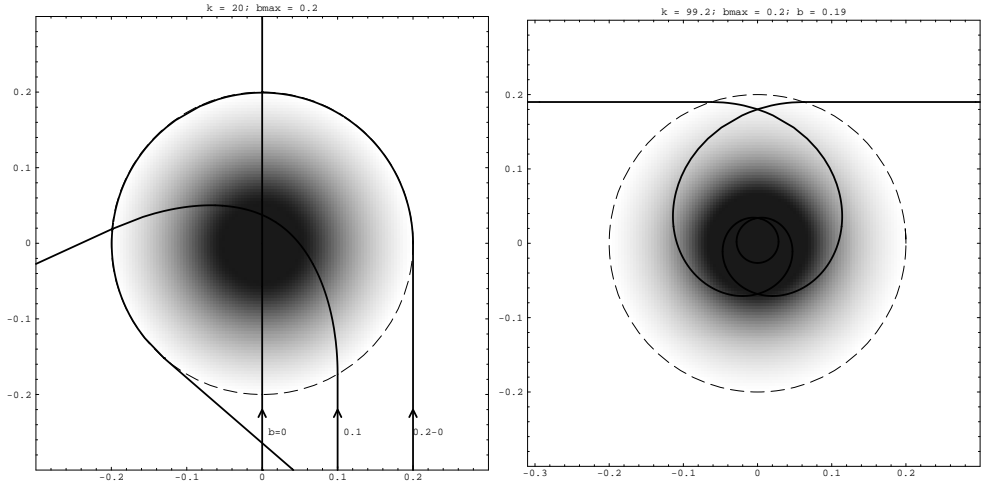

FIGURE 4. Some sample trajectories in the potential $V_{Z_K}$ before the background of a density plot of the potential. In the right figure the trajectory winds 3 times around the center.

(a,b) show some sample trajectories in the potential $V_{Z_K}$.

*Thus, the family $T_{k,b_{\max}}$ together with a parameter set $\mathcal{P} \subset \mathbb{R} \times ]0, 0.5[$ defines a 2-parameter family of potential scattering systems.* The associated family of trapped sets is

$$\begin{aligned}
\Lambda_{k,b_{\max}} &:= \Lambda(T_{k,b_{\max}}, Q_{b_{\max}}) \\
&= \{u = (x,b) \in Q_{b_{\max}};\ |\pi_p(T^i_{k,b_{\max}}(u))| \leq b_{\max}\ \forall\ i \in \mathbb{Z}\} \\
&= \{u \in \Lambda_{k,b'_{\max}};\ |\pi_b\left(T^i_k(u)\right)| \leq b_{\max}\ \forall\ i \in \mathbb{Z}\} \text{ for any } b'_{\max} > b_{\max}
\end{aligned}$$

where the pruning map $\pi_b$ is the projection onto the $b$-component. Varying only the parameter $k$ will lead to a deformation of a given trapped trajectory, which then may miss one of the unchanged scattering potentials and become a free trajectory (cf. figure 5 (a)). Varying only $b_{\max}$ changes the selection rule $|b| \leq b_{\max}$, so that a formerly trapped trajectory might miss one of the shrunken potentials and again become free (cf. figure 5 (b)). In both cases the trapped set changes. The difference between the two cases is that varying only $b_{\max}$ does not change the scattering map restricted to the shrunken interaction set, i.e. for $b^1_{\max} < b^2_{\max}$ we have $T_{k,b^2_{\max}}|Q_{b^1_{\max}} = T_{k,b^1_{\max}}|Q_{b^1_{\max}}$. Figure 6 shows the truncated sawshoe.

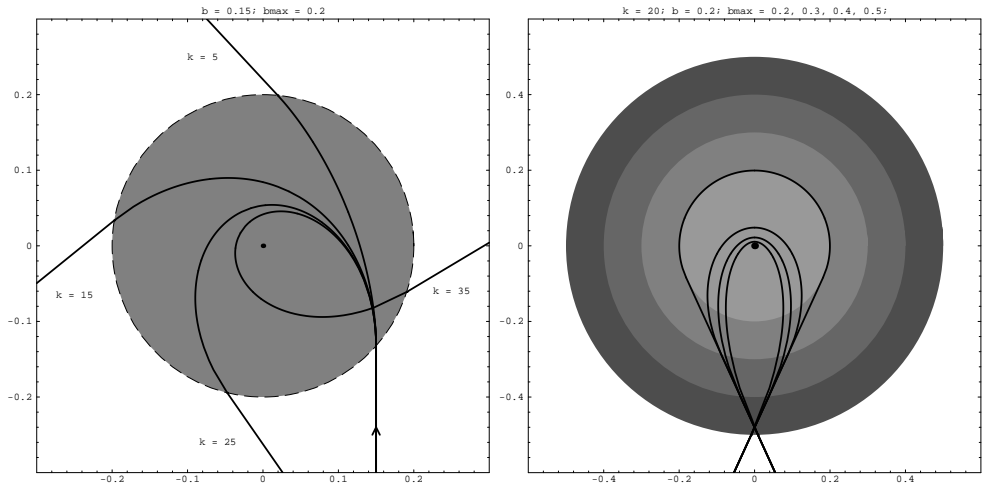

FIGURE 5. Example of (a) the influence of a variation of the parameter $b_{\max}$ on a trajectory with fixed impact parameter $b$ and (b) of a variation of the parameter $k$.

Now we shall introduce another simplification. From now on we restrict ourselves to the piecewise linearization of the scattering map given by

$$
(10) \qquad G(x) = \begin{cases} -x & \text{if } |x| < \pi/2 \\ -x + \pi & \text{if } |x - \pi| < \pi/2 \end{cases}
$$

This linearization corresponds to the limit $b_{\max} \to 0$ in the physical scattering family and describes therefore its asymptotic behaviour. In the parameter set $\mathcal{P}$ defined as

$$
(11) \qquad \mathcal{P} = \{(k, b_{\max}) \; ; \; 0 < b_{\max} < 0.5, k > 4, \pi \leq k\, b_{\max} < 2\pi\}
$$

we observe a transition from a regular to a chaotic dynamics via a pruning process, as will be made clearer presently.

A (generalized) symbolic dynamics for the sawtooth family which yields a simple symbolic form of the pruning map $\pi_b$ is the following:

*The sawtooth family is 2:1 semi-conjugate to the following pruned family over the alphabet $\mathring{A} = \{-1, 0, 1\}$:*

$$
(12) \qquad (\Sigma_\mu, \bar{\sigma}) \text{ with } \mu \in \mathring{B}^{\mathbb{N}}
$$

*whose dynamical map $\bar{\sigma} = -\sigma$ is the negative shift map (i.e. still a trivial cellular automaton). The trapped $\bar{\sigma}$-spaces $\Sigma_\mu = \Lambda(\bar{\sigma}, \Gamma_\mu)$ are determined by polynomial[6] pruning in the alphabet $\mathring{B} = \{-2, -1, 0, 1, 2\}$:*

$$
(13) \qquad \Gamma_\mu = \{a \in \mathring{A}^{\mathbb{Z}} : |P(a)| \leq \mu\}
$$
$$
P_0(a) = a_0; \; P_i(a) = a_{-i} + a_i, \; i \geq 1
$$

*with pruning map*

$$
(14) \qquad P(a) = \sum_{i=0}^{\infty} P_i(a) 5^{-i}, \quad P(\mathbf{a}) := \sup_{a \in \mathbf{a}} P(a)
$$

---

[6]It is called polynomial because the pruning functions $P_i$ are all polynomials.

FIGURE 6. The "sawshoe" $\Lambda_{k,b_{\max}}$ generated in phase space by the scattering system $(T_{k,b_{\max}}, Q_{b_{\max}})$ with interaction set $Q_{b_{\max}}$ shown in figure (a) and (b) is incomplete for the right figure (b) with the smaller $b_{\max}$ (and consequently smaller $Q_{b_{\max}}$).

Observe that we again interpret sequences with nonegative indices as real numbers, here in the base 5.

*Remark* 2.1. The truncation scenario of the sawshoe does not depend on the linearization (10). The nonlinear dynamical system determined by (7) and (8) is hyperbolic over an open parameter subset and undergoes there a hyperbolic transition to chaos by truncation of its horseshoe if one restricts to $k > 4$ and $b_{\max}$ small enough, because hyperbolicity is stable with respect to smooth perturbations and indeed the sine function converges uniformly toward its linearizations around its zeros for $b_{\max} \to 0$. One case of this claim is even trivial, namely if one fixes a $k > 4$ big enough and $b_{\max}$ small enough so that both the system is hyperbolic and the symbolic dynamics is the full shift $\mathring{A}^{\mathbb{Z}}$ and if subsequently one starts to decrease $b_{\max}$ down to zero. Obviously the hyperbolicity is conserved under such a parameter variation, because the scattering map changes only for the escaping points. Observe that trajectories change their form only within the potential's support (see figure 5) and will be pruned if and only if the potential's range shrinks below their largest impact parameter. The difference between the linearized and the nonlinearized system is that for the latter the pruning map $P$ is more complicated due to a loss of symmetry (cf. [18]).

## 3. Transition to Chaos of formal multiscattering systems

In this section we are going to introduce the concept of a formal multiscattering system[7] (FMS) in order to isolate the essential features of the examples given above. We shall give sufficient conditions for the existence of a SD for a FMS and we shall show how a FMS can become chaotic by truncation of the associated horseshoe. Finally we shall apply these concepts to the examples introduced above.

### 3.1. Formal Multiscattering systems (FMS).

**Definition 3.1.** Let $M$ be a metrizable space and $T : M \to M$ be a bijective map. Furthermore, let $Q \subset M$ be a compact subset and define the following images:

$$(15) \qquad I^i := T^i(Q) \cap Q, \quad i \in \mathbb{Z}$$

---

[7]In computer science (cf. [13]) a similar concept neglecting however topological structures is used and has recently been applied to the study of dynamical systems, namely the so-called *predicate transformers*.

The pair $(T, Q)$ is called a *formal multiscattering system* (FMS) if the following conditions are satisfied:

(i) **homeomorphism property:** The restriction $T|Q \to T(Q)$ is a homeomorphism[8].
(ii) **escape property:** For all $n \in \mathbb{N}$ : $T^{n+1}(I^{-n}) \setminus Q \neq \emptyset$.
(iii) **trapping property:** For all $n \in \mathbb{N}$ : $T^{n+1}(I^{-n}) \cap Q \neq \emptyset$.
(iv) **no-return property:** if $x \in Q$, $T(x) \notin Q$ then $\forall n \in \mathbb{N}$ : $T^n(x) \notin Q$.

From the definition one concludes that he trapped set $\Lambda$ of a FMS is non-empty and compact.

A sufficient property for the existence of a symbolic dynamics (SD) for a FMS is expansiveness:

**Definition 3.2.** (i) A FMS $(T, Q)$ is called *expansive* if $T|\Lambda$ is expansive in the usual sense, i.e. if there is a $\delta > 0$ s.t. $x \neq y$ implies the existence of a $n \in \mathbb{Z}$ with $d(T^n x, T^n y) > \delta$.
(ii) A finite open cover $\alpha$ of $\Lambda(T, Q)$ is called a generator for the FMS $(T, Q)$ if $\forall$ sequences $(A_i) \in \alpha^{\mathbb{Z}}$ : $\bigcap T^i(\bar{A}_i)$ contains no more than 1 point in $Q$. If $\bigcap T^i(A_i)$ contains no more than 1 point, $\alpha$ is called a weak generator.

Examples include the truncated double horseshoe and sawshoe from above (because they are hyperbolic) and the three disk scattering model because reflection at disks defocuses trajectories.

As in the standard situation of a bounded dynamics (cf. [20]) the existence of a generator is tantamount to expansiveness. As one can use a generator to construct a SD, expansiveness implies the existence of a SD. More precisely, one has the following three statements:

**Lemma 1.** *The FMS is expansive iff it has a generator iff it has a weak generator.*

**Theorem 3.1.** *If $(T, Q)$ is an expansive FMS, then there is a surjective semi-conjugacy $\Phi : \Sigma \to \Lambda(T, Q)$ from a subshift $\Sigma$ over some finite alphabet $\mathring{A}$.*

**Corollary 1.** *If the FMS $(T, Q)$ has a disjoint generator, then $\Phi$ can be chosen to be a topological conjugacy. In this case the trapped set $\Lambda(T, Q)$ is totally disconnected.*

---

[8]i.e. bijective and both $T$ and $T^{-1}$ are continuous

Finally we are going to make the notion of truncated families of FMS more precise:

**Definition 3.3.** A *truncated family of FMS* is given by a parameterized family of FMS $(T_\kappa, Q_\kappa)_{\kappa \in J}$ over some interval $J \subset \mathbb{R}^n$ as parameter set, where each $Q_\kappa$ is the truncation of a common set $Q \supset Q_\kappa$ by a family of level lines given by a cut-off function $f : M \to \mathbb{R}$ and an evaluation function $e : J \to \mathbb{R}^+$:

$$Q_\kappa = \{u \in Q : |f(u)| \leq e(\kappa)\} \tag{16}$$

**3.2. Transition to Chaos.** Expansiveness of a FMS is not quite enough for chaos because $\Lambda$ may contain isolated points:

**Definition 3.4.** Let $(T, Q)$ be a FMS with trapped set $\Lambda$. We call $(T, Q)$ *weakly chaotic* if its topological entropy[9] $H_{\text{top}} > 0$. The scattering dynamics is called *strongly chaotic* (or chaotic in the sense of Devaney) if (cf. [5]):

(i) the restricted map $T|\Lambda$ is sensitively dependent on initial conditions (s.d.i.c.).[10],
(ii) there is a dense orbit in $\Lambda$ and
(iii) the set of periodic points $\text{Per}_T(\Lambda)$ is dense in $\Lambda$:
$\Pi_T(\Lambda) := \overline{\text{Per}_T(\Lambda)} = \Lambda$.

In [3] it is shown that part (i) can be replaced by requiring $\Lambda$ to have infinitely many points.

We call the opposite of s.d.i.c. stability. If $T|\Lambda(T,Q)$ is stable everywhere on the trapped set, which means in this context that $\Lambda$ is discrete, then we speak of a *stable* or *regular* dynamics.

Whereas the symbolic dynamics of a full shift with number of symbols $\#\mathring{A} > 1$ is always strongly chaotic and that of a finite subshift is strongly chaotic iff the transition matrix is irreducible and aperiodic, there is no simple criterion for infinite subshifts.

The following questions about pruned multiscattering families and in particular about their symbolic models are of interest for us in this connection:

- Where is the map $\mu \mapsto \Sigma_\mu$ not locally constant, i.e. where does the trapped set acquire new points ? These values will be called *bifurcation values* forming the bifurcation set $\mathcal{V} = \mathcal{V}(\tau, \Gamma)$.

---

[9]As $\Lambda$ is compact, it is discrete iff it is finite. Hence the topological entropy $H_{\text{top}}(T,Q) := \lim_{r \to \infty} \frac{\log \#\{u \in \Lambda; \ \text{per}_T(u) \leq r\}}{r} > 0$ implies that $\Lambda$ cannot be discrete.

[10]In particular, $\Lambda$ cannot contain isolated points.

- Where does the scattering dynamics change its stability property? We call a bifurcation value $\nu_{\text{crit}} \in \mathcal{V}$ a *critical value*, if the stability property as function of $\mu$ is not locally constant at $\mu = \nu_{\text{crit}}$, or more precisely if for all neighbourhoods $U$ of $\nu_{\text{crit}}$ the dynamics $\tau|\Sigma_\mu$ is (weakly or strongly) chaotic for some $\mu = \mu_0 \in U$ and stable for some other $\mu = \mu_1 \in U$.
- How can $\Sigma_\nu$ be characterized at the bifurcation values $\nu \in \mathcal{V}$ and especially at the critical value(s)?

In the systems we are going to study the transitions are of the type where the restricted dynamics $\tau|\Sigma_\nu$ is stable for $\nu < \nu_{\text{crit}}$ and weakly chaotic in an open interval $(\nu_{\text{crit}}, \nu_{\text{crit}} + \epsilon)$. We call such a transition to chaos *abrupt* or *sudden* if $\tau|\Sigma_{\nu_{\text{crit}}}$ is stable, we call it a *cascade transition* if $\tau|\Sigma_{\nu_{\text{crit}}}$ is unstable, if furthermore the bifurcation set $\mathcal{V} \cap (-\infty, \nu_{\text{crit}})$ has only isolated points with the exception of one accumulation point $\nu_{\text{crit}}$ and if $\Sigma_{\nu_{\text{crit}}-\epsilon}$ is a finite set of cycles for all $\epsilon$ small enough. The notion of a cascade transition is of course motivated from similar transitions in the context of dissipative dynamical systems, e.g. the period-doubling cascade of the logistic map.

**3.3. Transition to chaos of the double horseshoe.** The bifurcation set of the double horseshoe family can be described as follows:

**Proposition 1.** *The subshifts $\Sigma_\mu = \{a \in \{0,1\}^{\mathbb{Z}} : P(\sigma^i(a)) \leq \mu \forall i \in \mathbb{Z}\}$ have the following specification:*

$$\Sigma_\mu = \{a \in \{0,1\}^{\mathbb{Z}} : (a_i)_{i \geq l} \leq \mu \ \forall l \in \mathbb{Z}\} \tag{17}$$

*The bifurcation set $\mathcal{V}(\sigma, \Gamma) \subset [0,2]$ can be characterized by a maximality property w.r.t. the 1-sided shift (Bernoulli-shift) $\sigma_B((\mu_i)_{i=0}^\infty) := (\mu_i)_{i=1}^\infty$:*

$$\mathcal{V}(\sigma, \Gamma) = \{\mu \in \{0,1\}^{\mathbb{N}_0}; \mu = \max_{k=0}^{\infty} \sigma_B^k(\mu)\} \tag{18}$$

*Proof.* The first statement is trivial. (i) If $\nu \in [1,2]$ has the form stated in the proposition, choose $\mathbf{a} = \bar{0}\nu$. We have $P(\mathbf{a}) = \nu$ because of the maximality property of $\nu$. Hence any $\Sigma_\mu$ with $\mu < \nu$ does not contain $\mathbf{a}$, so that $\Sigma_\nu$ is not local constant at $\mu = \nu$.

(ii) If $\mu < 1$ then symbol '1' is not allowed, $\Sigma_\mu/\sigma = \{\bar{0}\}$. If $\mu \geq 1$ does not have the maximality property then there is no $\mathbf{a} \in \Sigma_\mu/\sigma$ with $P(\mathbf{a}) = \mu$ because otherwise for an $a \in \mathbf{a}$ we would get $P(\sigma^{k_\delta}(a)) \geq \mu - \delta/2$ and $\sigma_B^{i_\delta}(\mu - \delta) \geq \mu + \delta$ for any $\delta > 0$ small enough, which implies $P(\sigma^{i_\delta + k_\delta}(a)) > \mu$. □

Another characterization of the bifurcation value is given by the topological entropy of the trapped set. Define a map $\gamma : [1, \infty[ \to \mathcal{V}(\sigma, \Gamma), \beta \mapsto a$ with

$$a_n := [\beta^n(1 - \sum_{i=1}^{n-1} a_i \beta^{-i})] \tag{19}$$

where $n \geq 1$ and $[x]$ denotes the largest integer strictly smaller than $x$. Then one has:

**Proposition 2.** (1) $\gamma$ is bijective and monotone increasing.
   (2) $\gamma^{-1}$ is continuous (but not $\gamma$).
   (3) $\log \gamma^{-1}(a) = H_{top}(\sigma|_{\Sigma_a})$.
   (4) $a \mapsto H_{top}(\sigma|_{\Sigma_a})$ is a Cantor function (devil's staircase).

*Proof.* Part (3) is shown in [20] by using a renewal theorem. The rest can be concluded from the fact that the following three assertions are equivalent:

   (i) $a = \gamma(\beta)$
   (ii) For all $i \geq 1$: $0 \leq a_i \leq [\beta]$ and $\sum_{i=1}^{\infty} a_i \beta^{-i} = 1$ and $\sigma_B^m(a) \leq a$ for all $m \geq 0$.
   (iii) for all $m \geq 1$: $1 - \beta^{-m} \leq \sum_{i=1}^{m} a_i \beta^{-i} \leq 1$.
□

The onset of chaos can now be described as:

**Corollary 2.** *The bifurcation set $\mathcal{V}(\sigma, \Gamma)$ consists of a discrete part $\mathcal{V} \cap [0, 1] = \{0\}$ and a Cantor set $\mathcal{V} \cap [1, 2]$. The bifurcation values $\nu \in \mathcal{V}$ can be classified according to the number of consecutive '1's allowed in the corresponding $\Sigma_\nu$:*

| | |
|---|---|
| $\mu < 1$ | $\mu \notin \mathcal{V}$; $\Sigma_\mu/\sigma = \{\bar{0}\}$: no symbol '1' allowed |
| $\mu = 1$ | $1 \in \mathcal{V}$; $\Sigma_1/\sigma = \{\bar{0}, \bar{0}1\bar{0}\}$: first appearance of (isolated) symbol '1'; |
| $1 \leq \mu < 1.1$ | $\mu \in \mathcal{V} \Rightarrow \mu = 1.0^{r_0} 1 0^{r_1} \dots 1 0^{r_i} \dots$ with $r = (r_i)_{i=0}^\infty = \min\{(r_i)_{i>j} : j \in \mathbb{N}_0\}$; |
| $\mu = 1.1$ | $1.1 \in \mathcal{V}$; first appearance of a pair '11'; |
| etc. | |
| $\mu = 1.\bar{1} = 2$ | $1.\bar{1} \in \mathcal{V}$; $\Sigma_{1.\bar{1}} = \{0, 1\}^\mathbb{Z}$: full shift. |

*Proof.* $\mathcal{V} \cap [1, 2]$ is disconnected by numbers whose binary representation contains a segment $1^r$ with $r > 0$ large enough. Its complement is obviously open, and by proposition 1 none of its points is isolated. □

**Corollary 3.** *The onset of chaos of the family* $(\sigma, \Sigma_\mu)$ *is sudden at* $\nu_{crit} = 1$ *and for all* $\mu > \nu_{crit}$ *the topological entropy* $H_{top} > 0$ *and the dynamics* $(\sigma, \Sigma_\mu)$ *is strongly chaotic.*

*Proof.* All points in $\Sigma_\mu$, $\mu > 1$, are accumulation points, because one can exchange the tail of any $a \in \Sigma_\mu$ by $0^r 1\bar{0}$ for $r > 0$ large enough. The periodic sequences are dense because one can imitate any $a \in \Sigma_\mu$ up to any indices $n > 0, m < 0$, then add a segment $0^r 10^r$, $r > 0$ large enough, and repeat the whole periodically. A dense orbit is constructed by segments of periodic sequences which are connected by segments $0^r 10^r$, $r > 0$ large enough. $\square$

### 3.4. Transition to chaos of the sawshoe.
The corresponding study of the more complicated sawshoe yields:

**Theorem 3.2 (Troll 91).** *The transition to chaos of the sawshoe family occurs via a cascade, whose bifurcation values in symbolic parameter space are*

$$\pi^{(-2)} := 0 \cdot \bar{0},$$
$$\pi^{(-1)} := 1 \cdot \overline{\text{-}22},$$
$$\pi^{(n)} := 1 \cdot \overline{\text{-}10^n\text{-}12} \quad \text{for } n \geq 0,$$
$$\pi^{(\infty)} := 1 \cdot \text{-}1\bar{0} \tag{20}$$

*and which accumulates in the real parameter space towards the critical line*

$$\partial_o \mathcal{P}_0 = \{(k, b_{\max}) \in \mathcal{P};\ k > 7.2 \text{ and } k\, b_{\max} = \frac{\pi^2}{\pi - b_{\max}}\} \tag{21}$$

*The dynamical systems on the critical line are unstable with* $H_{top} = 0$ *and the cycles are dense in the set of trapped orbits.*

To characterize the transition in terms of structural complexity we need the following representation of the periodic regime of the dynamics:

**Proposition 3 (Troll 91).** *Each cycle* $\mathbf{a} \in \text{Per}(\Sigma_{\pi^\infty})/\bar{\sigma}$ *can be generated from the cycles* $\bar{0}$ *and* $\overline{\text{-}11}$ *by the transformation semi-group spanned by the following maps operating on* $\Sigma_{\pi^\infty}$:

(i) <u>reflection map:</u>

$$t_R(\mathbf{a}) := -\mathbf{a} \tag{22}$$

(ii) <u>basic inflations</u> *operating as replacement rules for single symbols :*

(23)
$$I^+_{[1]} : \quad \begin{array}{rcl} -1, 0 & \mapsto & 0\text{-}11 \\ 1 & \mapsto & 0^2\text{-}11 \end{array} \qquad I^-_{[1]} : \quad \begin{array}{rcl} -1 & \mapsto & 0\text{-}11 \\ 0, 1 & \mapsto & 0^2\text{-}11 \end{array}$$

$$I^+_{[2]} : \quad \begin{array}{rcl} -1, 0 & \mapsto & 0\text{-}11 \\ 1 & \mapsto & 0(\text{-}11)^2 \end{array} \qquad I^-_{[2]} : \quad \begin{array}{rcl} -1 & \mapsto & 0\text{-}11 \\ 0, 1 & \mapsto & 0(\text{-}11)^2 \end{array}$$

(iii) <u>repetition map</u>
Each $\mathbf{a} \in \text{Per}(\Sigma_{\pi^\infty})$, $\mathbf{a} \neq \bar{0}$, can be formed by concatenating blocks $B_{r[\alpha]}$ with fixed block class $\alpha \in \{1, 2\}$ and varying block order $r \in \mathbb{Z} \setminus \{0\}$, where $B_{|r|[1]} := 0^{|r|}\text{-}11$, $B_{|r|[2]} := 0(\text{-}11)^{|r|}$ and $B_{-|r|[\alpha]} := -B_{|r|[\alpha]}$.
The repetition map $A$ operates now on the constituent blocks $B_{r[\alpha]}$ as follows:

$$A : B_{r[\alpha]} \mapsto B_{r+\text{sgn}(r)[\alpha]} \tag{24}$$

For $\mathbf{a} = \bar{0}$ we define $A(\bar{0}) := \bar{0}$.

The way orbits are generated from the fixed point $\bar{0}$ by this semi-group is illustrated in figure 7.

## 4. Complexity : Subshifts and Languages

The problem addressed in this section is how to characterize the transition to chaos of pruned systems. The onset of chaos in such systems indicates features of a phase transition, seen for instance in the topological entropy as a function of the family parameters (cf. proposition 2 and [16]). It was conjectured (cf. [4]) that high level computation occurs at phase transitions. One indication is the appearance of infinite subshifts as a consequence of the divergence of a "grammatical" correlation length, which specifies how many predecessors of a symbol must be known to determine whether it is allowed or forbidden. It seems to be natural to interpret such a singularity at a critical point as indicating a complexity jump which requires a new and more complex production "mode" of the pruned SD. The theory of automata and formal languages offers concepts to make this more precise. Appendix C.1 summarizes some basic notions and notations. But first we have to show how to connect the (infinite) sequences of a SD with the finite words of a formal language.

FIGURE 7. Example of the generation of a cycle by inflation. The codes of the cycles are given both in a 3-symbol generalized 2:1 SD and a 6-symbol 1:1 SD. The centers of the potentials are marked by cogwheel symbols.

**4.1. Dynamically generated languages.** Here we identify those languages which are generated by subshifts. These are the associated *central* language $\mathcal{L}(\Sigma)$ which consists of all non-empty segments of permitted sequences and the associated *cyclic* language $\mathcal{L}_c(\Sigma)$, which consists of all full periods of a cycle. Observe that $\mathcal{L}_c(\Sigma)$ contains together with any word $w$ all its repetitions $w^i$, $i > 0$ and all their cyclic permutations.

An important property of these dynamically generated languages is that $\mathcal{L}(\Sigma)$ determines $\Sigma$ uniquely and so does $\mathcal{L}_c(\Sigma)$, provided $\Sigma$ is a cyclic subshift (def. cf. appendix). Therefore different subshifts can still be distinguished on the level of their associated languages. It is possible to describe all dynamically generated languages on the language level. For that purpose one introduces the center and the cyclic center of a language:

(i) The *center* $\mathcal{C}(L)$ of a language $L$ consists of all biextensible words:

$$(25) \quad \mathcal{C}(L) = \{a \in \mathring{A}^+; \text{for all } N > 0 \text{ there are words } x, y \in \mathring{A}^*$$
$$\text{of lengths } \lg(x), \lg(y) \geq N \text{ s.t. } xay \in L\}$$

(ii) The *cyclic center* $\mathcal{C}_c(L)$ is

$$(26) \quad \mathcal{C}_c(L) = \{a \in \mathring{A}^+; \forall \text{ repetitions } i > 0 \; \forall \text{ cyclic permutations } \pi(a^i)$$
$$\text{there are strings } x, y \in \mathring{A}^* \text{ s.t. } x\pi(a^i)y \in L\}$$

We call a language $L$ *central* if $\mathcal{C}(L) = L$ and *cyclic* if $\mathcal{C}_c(L) = L$. One sees from this definition easily that the associated central language of a subshift is central and that the associated cyclic language of a cyclic subshift is cyclic. The converse statement is also true, so that

**Proposition 4.** *There is a bijective correspondence between subshifts and central languages and between cyclic subshifts and cyclic languages.*

**4.2. Complexity Transition.** We measure the structural complexity of any language $L$ by its minimal computational model, in particular by the simplest[11] Chomsky class it belongs to. We define the Chomsky complexity $\chi(L)$ of any language $L$ as (-1) times[12] the type number of the simplest machine accepting it. A complexity transition is then roughly a change of the Chomsky complexity. However, we prefer to consider a particular symbolic dynamics only as a realization of a given discrete dynamical system. There are many other equivalent symbolic dynamics. We choose a particular one among them by convenience in much the same way as we choose a convenient coordinate system for a geometric problem. Therefore, we have to make sure that a local complexity change occurs in each equivalent family of symbolic dynamics. Indeed, this independence of the symbolic dynamics is a general theorem (cf. [17, 19]).

---

[11]Observe that we chose the Chomksy type number to increase with decreasing complexity.
[12]because we prefer the reversed order;

**Definition 4.1.** The pruned system $(\tau_\mu, \Sigma_\mu)_{\mu \in J}$ shows a *complexity transition* at the critical value $\mu_{\text{crit}} \in J$ if for any equivalent symbolic dynamics $(\sigma, \tilde{\Sigma}_\mu)_{\mu \in J}$, the Chomsky type function $\kappa \mapsto \chi(\mathcal{L}(\Sigma_\mu))$ is not locally constant at $\mu = \mu_{\text{crit}}$, i.e. assumes more than one value in any neighborhood of $\mu_{\text{crit}}$.

**Theorem 4.1 (Troll 92).** *Any semi-conjugacy $\Phi$ between two subshifts does not increase the Chomsky complexity of the associated (cyclic or central) language. In particular, if $\Phi$ is even a conjugacy then:*

$$\chi(\mathcal{L}(\Sigma_\mathfrak{A})) = \chi(\mathcal{L}(\Sigma_\mathfrak{B})) \tag{27}$$

$$\chi(\mathcal{L}_c(\Sigma_\mathfrak{A})) = \chi(\mathcal{L}_c(\Sigma_\mathfrak{B})) \tag{28}$$

*where $\Sigma_\mathfrak{B} = \Phi(\Sigma_\mathfrak{A})$ and we assumed for the second equality that $\Sigma_\mathfrak{A}$ is cyclic.*

**4.3. Complexity transitions in truncated horseshoes.** We state the main results for our horseshoe models. First for the double horseshoe:

**Theorem 4.2 (Troll 92).** *Suppose $\nu \in \mathcal{V} \cap ]1, 2]$. Then both the associated central language $\mathcal{L}(\Sigma_\mu)$ and the cyclic language $\mathcal{L}_c(\Sigma_\mu)$ are regular languages if and only if $\nu \in ]1, 2]$ is rational. The associated languages are finite complement languages (i.e. $\Sigma_\mu$ is a finite subshift) iff $\nu$ has a periodic binary representation without leading transient part.*

**Corollary 4.** *A complexity transition occurs for the truncated double horseshoe simultaneously with the onset of chaos at $\nu_{crit} = 1$.*

For the sawshoe one can state:

**Theorem 4.3 (Troll 91).** *The parameter family of the associated formal (either central or cyclic) languages $\mathcal{L}_\mu$ has the following properties:*

(a) *in $\mathcal{P}_0 = \{(k, b_{\max}) \in \mathcal{P}; \, k > 7.2 \text{ and } k\, b_{\max} < \frac{\pi^2}{\pi - b_{\max}}\}$ the language $\mathcal{L}_\mu$ is finite (hence regular);*

(b) *on $\partial_o \mathcal{P}_0$ the language $\mathcal{L}_\mu$ is not context free but context sensitive.*

æ

## Appendix A. Basic definitions in symbolic dynamics

We discuss here dynamical systems whose dynamical map operates on sequences. We permit sequences over some finite set $\mathring{A}$ of abstract symbols with cardinality $\#\mathring{A} > 1$ and define the *full shift* over $\mathring{A}$ as the complete set of sequences $\mathring{A}^{\mathbb{Z}}$.

We can make a metric space out of $\mathring{A}^{\mathbb{Z}}$ by defining distances between sequences:

$$(29) \qquad d(a,b) = \sum_{i \in \mathbb{Z}} \frac{\bar{\delta}_{a_i b_i}}{2^{|i|}}$$

where $a, b \in \mathring{A}^{\mathbb{Z}}$, $\bar{\delta}_{xy} = 0$ if $x = y$, 1 otherwise.

Among the continuous maps operating on $\mathring{A}^{\mathbb{Z}}$ shift maps and cellular automata (CA) are of special interest because they are compatible with the group structure of the index set $\mathbb{Z}$:

Denote by $\sigma_{\mathring{A}}$ the (right) *shift map* on $\mathring{A}^{\mathbb{Z}}$: $(\sigma(a))_i = a_{i+1}$. Usually we write just $\sigma$ for $\sigma_{\mathring{A}}$. A *subshift* is a closed $\sigma$-invariant set in $\mathring{A}^{\mathbb{Z}}$. $\Sigma$ is a *finite subshift* if it is determined by a finite set $F$ of forbidden symbol strings, which are not permitted to appear in sequences. A subshift is *Markov* if the immediate predecessor determines whether a symbol is allowed or forbidden. Such a Markov rule is usually given in the form of a transition matrix indexed by the alphabet and showing an entry 1 wherever a pair of symbols is allowed and 0 otherwise. We call $\Sigma$ *cyclic* if the periodic (w.r.t. $\sigma$) sequences are dense. A dynamical system given by s subshift $\Sigma \subset \mathring{A}^{\mathbb{Z}}$ and the shift map $\sigma|\Sigma \to \Sigma$ is called a *symbolic dynamics*.

A natural generalization leads to *cellular automata (CA)*. A (1-dimensional) CA is any continuous map $\tau$ on $\mathring{A}^{\mathbb{Z}}$ which commutes with the shift map.[13]

A way to compare dynamical systems topologically is offered by *semi-conjugacies*. We define them here just for shift maps as any continuous map $\Phi$ between subshifts $\Sigma_{\mathring{A}}$, $\Sigma_{\mathring{B}}$ which intertwines between the respective shift maps $\sigma_{\mathring{A}}$, $\sigma_{\mathring{B}}$:

$$(30) \qquad \Phi : \Sigma_{\mathring{A}} \to \Sigma_{\mathring{B}}, \quad \sigma_{\mathring{B}} \circ \Phi = \Phi \circ \sigma_{\mathring{A}}$$

If moreover, $\Phi$ is a homeomorphism i.e. bijective and also $\Phi^{-1}$ continuous, then it is called a *conjugacy* (between shift maps). It may be interpreted as a continuous change of variables which identifies the shift maps on $\Sigma_{\mathring{A}}$ and $\Sigma_{\mathring{B}}$. For instance, finiteness and the Markov property of subshifts are equivalent in this sense, i.e a subshift $\Sigma$ is finite iff it is topologically conjugate to a Markov subshift.

---

[13]Since any CA is automatically uniformly continuous an equivalent description of a CA (cf. [8, 9]) can be given by a local function $f_\tau : \mathring{A}^{2R+1} \to \mathring{A}$ with the property $(\tau(s))_i = f_\tau(s_{i-R}, \ldots, s_{i+R})$ for any $s \in \mathring{A}^{\mathbb{Z}}$.

## Appendix B. The inverse scattering problem for the sawtooth deflection

**B.1. Statement of the inverse problem.** We are looking for an attractive central potential $V\ :\,]0,\infty[\to\mathbb{R}$ whose deflection function $\theta_V$ on $]0,b_{\max}[\cup]b_{\max},\infty[$ (allowing for singularities at $0$ and $b_{\max}$) for a given asymptotic particle energy $E_0>0$ is the sawtooth $Z_k$, $k>0$. We require the following properties:

(1) $V\leq 0$, $V'\geq 0$
(2) $V(r)=0$ if $r\geq b_{\max}$,
(3) $V$ continuous,
(4) $V$ smooth ($C^2$ suffices) in $]0,b_{\max}[$.

The forward problem of finding the deflection function $\theta_V$ of a given (not necessarily attractive) central potential $V$ of range $b_{\max}$ and a given asymptotic particle energy $E_0>0$ can be easily solved from the Hamilton equations to yield for impact parameter $0<b<b_{\max}$ the possibly infinite value:

$$\begin{aligned}(31)\quad \theta_V(b)&=2\int_{r_{\min}(b)}^{\infty}\frac{dr}{r^2\sqrt{(1-\frac{V(r)}{E_0})b^{-2}-r^{-2}}}-\pi\\ &=-2\arccos(\frac{b}{b_{\max}})+2\int_{r_{\min}(b)}^{b_{\max}}\frac{dr}{r^2\sqrt{(1-\frac{V(r)}{E_0})b^{-2}-r^{-2}}}\end{aligned}$$

where $r_{\min}(b)$ is the classical turning point (minimal $r$ value) of the trajectory. If $V'(r)=0$ (zero force) can be excluded, $r_{\min}(b)$ coincides with the most remote point satisfying $\dot r=0$, i.e. it is determined as the largest zero of the radicand $(1-\frac{V(r)}{E_0})b^{-2}-r^{-2}$:

$$(32)\qquad r_{\min}(b)=\max\{r\geq 0;\ 1-V(r)/E_0=b^2/r^2\}$$

For $b<0$ we set $\theta_V(-b):=-\theta_V(b)$.

**B.2. Solution of the inverse problem.** In order to present the solution formula conveniently we define a function $\rho$:

$$(33)\qquad \rho\ :\ [0,b_{\max}]\ \to\ [0,b_{\max}],\quad s\mapsto s\,e^{-k/\pi\sqrt{b_{\max}^2-s^2}}$$

As one easily verifies (see proof below), it is strictly monotone increasing, surjective, continuously differentiable and it has no critical points in $[0, b_{\max}[$. Hence, $\rho$ has a continuously differentiable inverse function $\xi$.

**Theorem B.1.** *A solution of the inverse scattering problem B.1 is given by:*

(34)
$$V_{Z_K} : [0, \infty[ \to \mathbb{R}_0^-, \quad r \mapsto \begin{cases} E_0 \left(1 - e^{2k/\pi \sqrt{b_{\max}^2 - \xi^2(r)}}\right) & \text{if } 0 \leq r \leq b_{\max} \\ 0 & \text{if } r > b_{\max} \end{cases}$$

*The solution is unique in the class of potentials satisfying* $V(r) \geq 1 - b_{\max}^2/r^2 \; \forall \, r \in \, ]0, b_{\max}[$.

A parameter representation of this potential is given by

(35)
$$\left\{ \left( s \, e^{-k/\pi \sqrt{b_{\max}^2 - s^2}}, \, E_0(1 - e^{2k/\pi \sqrt{b_{\max}^2 - s^2}}) \right); \, s \in [0, b_{\max}] \right\}$$

In the proof we shall proceed as follows. First we shall show that equation (34) solves the inverse problem B.1. Afterwards, we shall prove uniqueness.

B.2.1. *Existence.*

*Proof.* That $\rho$ is strictly monotone increasing and has no critical points in $[0, b_{\max}[$ can be seen by differentiating equation (33):

(36)
$$\rho'(s) = \frac{k \, s^2 + \pi \sqrt{b_{\max}^2 - s^2}}{e^{k/\pi \sqrt{b_{\max}^2 - s^2}} \, \pi \sqrt{b_{\max}^2 - s^2}}$$

Hence $\rho$ and its inverse $\xi$ are admissible integral substitutions in equation (31). Indeed they are even analytic functions, so that $V_{Z_K}$ is also analytic in $]0, b_{\max}[$ and, obviously, continuous in $[0, \infty[$. Substituting $s = \xi(r)$ and (34) in (31) yields

(37)
$$\theta(b) = k \, b$$

for $0 \leq b < b_{\max}$. $\square$

B.2.2. *Uniqueness.*

*Proof.* Let $V$ be a solution of the inverse problem B.1. We first observe that $\theta_V$ is most naturally a function of $b^{-2}$. Therefore we introduce

$$(38) \qquad y := b^{-2}, \quad y_0 := b_{\max}^2, \quad \hat{\theta}(y) := \theta_V(b), \quad \text{and} \quad v(r) := 1 - \frac{V(r)}{E_0}$$

Next we make a substitution in the integral of equation (31), whose purpose is to remove the implicit dependence of the integral limits (i.e. $r_{\min}$) on the potential $V$:

$$(39) \qquad x(r) := \frac{1}{r^2 \, v(r)}$$

**Lemma 2.** *If $v(r) \leq b_{\max}^2 / r^2 \, \forall \, r \in \,]0, b_{\max}[$, then the variable transformation $x$ is a positive and monotone function bounded by $r^{-2}$.*

*Proof.* The critical points of this transformation satisfy:

$$(40) \qquad x'(r) = -\frac{2r \, v(r) + r^2 v'(r)}{r^4 v^2(r)} = 0$$

i.e. $2r \, v(r) + r^2 v'(r) = 0$. The existence of a critical point $r_0$ is obviously equivalent to the existence of a constant $C > 0$ s.t. $v(r_0) = C/r_0^2$ and $v'(r_0) = -2\,C/r_0^3$, i.e. $v$ is locally inverse quadratic at any critical point. In order to show that equation (39) defines an a monotone function it suffices to show that $x$ does not have a critical point in the open interval $]0, b_{\max}[$.

**Lemma 3.** *Let $0 \leq b_0 \leq b_{\max}$ be the impact parameter of a trajectory with classical turning point $0 < r_0 < b_{\max}$, where the potential $v$ is assumed to be locally inverse quadratic. Then the deflection function of $v$ has a singularity at $b_0$ and is consequently unbounded.*

*Proof.* One way to see this is to set the Taylor expansion of $v$ around $r_0$ into (31). One finds a nonintegrable singularity $(r - r_0)^{-1}$. A more physical argument is given by the observation that criticality at $r_0$ is equivalent to the equality of the potential's gradient at $r_0$ and the centrifugal force for curvature radius $r_0$. Together with the turning point condition $\dot{r}(r_0) = 0$ and the uniqueness of the initial value problem of motion in a potential field, this yields a trajectory which ends as circular motion, winding infinitely often around the potential's center. □

Now, consider the turning point function

(41) $$r_{\min} : [0, b_{\max}] \to [0, b_{\max}], \quad b \mapsto \max\{r;\, r^2\, v(r) - b^2 = 0\}$$

Although $r_{\min}(0) = 0$ and $r_{\min}(b_{\max}) = b_{\max}$, not every critical point $r_0$ need to appear as a turning point for some impact parameter $b$, because the turning point function need not be continuous. However, since $v$ was supposed to be $C^2$ in $]0, b_{\max}[$, the only way a discontinuity may arise is via a tangent bifurcation at a local extremum of

(42) $$\hat{v} : ]0, \infty[ \to [0, \infty[, \quad r \mapsto r^2\, v(r)$$

Two cases of discontinuities of the turning point function $r_{\min}$ can be distinguished:

> **case 1:** There is a critical point at which $\hat{v}$ assumes a value $< b_{\max}^2$. But then there must be relative minima among the critical points. Since $\hat{v}$ is $C^2$ there is a lowest minimum. By construction it is both a critical point and a turning point. Thus the conditions of lemma 3 are satisfied, so that trajectories with infinite total curvature occur, in other words the deflection function must be unbounded. However, the sawtooth deflection function is bounded. Hence, in this case the integral substitution $x$ of (39) is a monotone function and consequently admissible.
>
> **borderline case:** Suppose there is a critical point of $\hat{v}$ on the line $b_{\max}^2$ that is the accumulation point of a sequence of turning points. This would again lead to a limit circle, so that this case is also excluded because $\theta_V$ would be unbounded.
>
> **case 2:** $\hat{v}$ assumes values $> b_{\max}$ and all critical points of $\hat{v}$ in $]0, b_{\max}[$ assume values $\geq b_{\max}^2$. In this case there is a $0 < r^* < b_{\max}$ s.t. $r^2\, v(r) \geq b_{\max}^2\ \forall\, b_{\max} > r \geq r^*$.

Positivity of $x$ is immediate, because we assumed $V$ to be nonpositive, so that $v(r) \geq 1$. □

Introducing the function

(43) $$g(x) = -\frac{\sqrt{x}}{r} \frac{dr(x)}{dx}$$

we get

(44) $$\hat{\theta}(y) = -2 \arccos\left(\frac{\sqrt{y_0}}{\sqrt{y}}\right) + 2 \int_{y_0}^{y} dx\, \frac{g(x)}{\sqrt{y - x}}$$

The important point is that $y = b^{-2}$ is both a limit of integration in the expression for the deflection function and a natural choice of argument for the deflection function itself. This makes it possible to write our inverse problem as an Abel integral equation. We first write the sawtooth as a function of $y$:

$$\hat{Z}_k(y) = \frac{k}{\sqrt{y}} \tag{45}$$

so that the integral with Abel kernel $(y-x)^{-1/2}$ and upper limit $y$ equals a function $\hat{\hat{\theta}}$ of $y$:

$$\hat{\hat{\theta}}(y) := \frac{k}{2\sqrt{y}} + \arccos\left(\frac{\sqrt{y_0}}{\sqrt{y}}\right) = \int_{y_0}^{y} dx \, \frac{g(x)}{\sqrt{y-x}} \tag{46}$$

This is an Abel integral equation for $g$ which can be solved by the following inversion formula, because $\hat{\hat{\theta}}$ is smooth in $]0, b_{\max}[$:

$$g(x) = \frac{\sin(\frac{\pi}{2})}{\pi} \left( \frac{\hat{\hat{\theta}}(y_0)}{(x-y_0)^{\frac{1}{2}}} + \int_{y_0}^{x} \frac{d\hat{\hat{\theta}}(y)}{dy} \frac{dy}{(x-y)^{\frac{1}{2}}} \right) \tag{47}$$

This yields

$$g(x) = \frac{1}{2\sqrt{x}} + \frac{k}{2\pi\sqrt{x-y_0}\sqrt{y_0}} - \frac{k\sqrt{x-y_0}}{2\pi x\sqrt{y_0}} \tag{48}$$

Now, in order to obtain $v$ one uses equations (39) and (43) to find the following expression for $w(x) := v(r(x))$:

$$\begin{aligned}\frac{d\ln(w(x))}{dx} &= -\frac{1}{x} + \frac{2g(x)}{\sqrt{x}} \\ &= \frac{k\sqrt{y_0}}{\pi x^{\frac{3}{2}}\sqrt{x-y_0}}\end{aligned} \tag{49}$$

This equation has to be integrated. We use $w(y_0) = v(b_{\max}) = 1$ (because $x(b_{\max}) = y_0$) to get:

$$\ln(w(x))) = \frac{2k\sqrt{x-y_0}}{\pi\sqrt{x}\sqrt{y_0}} \tag{50}$$

Hence

$$w(x) = e^{2k/\pi \sqrt{y_0^{-1} - x^{-1}}} \tag{51}$$

Now, we are ready to take the last step in calculating the wanted potential. Observe that

$$r(x) = \frac{1}{\sqrt{xw(x)}} = x^{-1/2} e^{-k/\pi \sqrt{y_0^{-1} - x^{-1}}} \tag{52}$$

is the inverse of $x(r)$. Although it is not possible to solve equation (52) explicitly for $x$, we can easily give a parameterization of the potential curve $(r, V_{Z_K}(r))$, $r \in \mathrm{supp}(V)$, in terms of $x \in [y_0, \infty[$:

$$\begin{aligned}(r(x), V_{Z_K}(r(x))) &= (r(x), E_0(1 - w(x))) \\ &= \left(x^{-1/2} e^{-k/\pi \sqrt{y_0^{-1} - x^{-1}}}, E_0(1 - e^{2k/\pi \sqrt{y_0^{-1} - x^{-1}}})\right)\end{aligned} \tag{53}$$

Monotone reparameterization by $s = x^{-1/2}$ yields the assertion. $\square$

Figure 3 shows this potential, which has the following properties:

**B.3. Properties of the potentials $V_{Z_K}$.** Each $V_{Z_K}$ is continuous at $b_{\max}$, because $w(y_0) = 1$ (see above). At $r = 0$, which corresponds to $x \to \infty$, one finds:

$$V_{\min} = V_{Z_K}(0) = E_0 \left(1 - e^{2 k b_{\max}/\pi}\right) \tag{54}$$

The derivative of $V_{Z_K}$ is given by

$$\left(r(s), V'_{Z_K}(r(s))\right) = \left(r(s), E_0 \frac{2 e^{3k/\pi \sqrt{b_{\max}^2 - s^2}} \, k\, s}{k\, s^2 + \pi \sqrt{b_{\max}^2 - s^2}}\right) \tag{55}$$

One sees from this expression that the potential $V_{Z_K}$ has independently of $k$ and $E_0$ a unique critical point ($V'_{Z_K} = 0$) at the center $r = s = 0$. At the border of the potential's support ($r = s = b_{\max}$) the left derivative is also independent of $k$, namely $V'_{Z_K}(b_{\max}) = 2E_0/b_{\max} \neq 0$. This is necessary to get a non vanishing (indeed the maximal) deflection just at the border of the potential.

The influence of a variation of the parameters $k$ and $b_{\max}$ on the shape of the potential $V_{Z_K}$ can be studied through the second derivative of $V_{Z_K}$:

$$(56) \quad (r(s), V''_{Z_K}(r(s))) = \left( r(s), E_0 \frac{2 e^{4k/\pi \sqrt{b_{\max}^2 - s^2}} k \left( b_{\max}^2 \pi^2 - 3 k^2 s^4 - 4 k \pi s^2 \sqrt{b_{\max}^2 - s^2} \right)}{\left( k s^2 + \pi \sqrt{b_{\max}^2 - s^2} \right)^3} \right)$$

At $r = 0$ one finds $V''_{Z_K}(0) = 2kE_0/(\pi b_{\max}) \exp(4kb_{\max}/\pi) \neq 0$. The zeros of $V''_{Z_K}$ are determined by the equation

$$(57) \quad \pi^2 b_{\max}^2 - 3 k^2 s^4 - 4 k \pi s^2 (b_{\max}^2 - s^2)^{1/2} = 0$$

The left hand side of this equation defines a function $f(s)$ with $f(0) = \pi^2 b_{\max}^2 > 0$. If $k > 0$ is too small then $f(s) > 0$ for all $0 \leq s \leq b_{\max} \leq 1/2$ and $V''_{Z_K}$ does not have any zeros. If however $f(b_{\max}) = (\pi^2 - 3k^2 b_{\max}^2) b_{\max}^2 \leq 0$, i.e. if

$$(58) \quad k\, b_{\max} \geq \frac{\pi}{\sqrt{3}}$$

then $V''_{Z_K}$ has zeros in the interval $[0, b_{\max}]$. For the third derivative one finds again a zero at $r = 0$: $V'''_{Z_K}(0) = 0$. The graphs of $V'_{Z_K}$ for some typical $k$-values are shown in figure 8.

**B.4. The unrestricted solution space of the inverse problem.** If one drops the condition that $\hat{v}$ must not have only critical points above $b_{\max}^2$ then the solution of the inverse problem is no longer unique. From the former sections we know that any solution cannot have a critical point below $b_{\max}^2$, because otherwise the deflection function would be unbounded. Let $r^* = \inf\{0 \leq r \leq b_{\max}; \hat{v}(r) = b_{\max}^2\}$. Then we can perform the integral transformation (39) in the interval $[0, r^*]$. The value $r^*$ and the potential in $[r^*, b_{\max}]$ can be chosen freely $\geq b_{\max}^2/r^2$ but in the neighbourhood of $r^*$ where we require smooth matching and at $b_{\max}$ where we require continuity. Going through the calculations above one finds that all solutions have the general form :

$$(59) \quad w^{\mathrm{gen}}(x) = \frac{b_{\max}^2}{r^{*2}} w(x) e^{2/\pi \int_{y_0}^{x} R_{r*}(y) \frac{d}{dy}(\mathrm{arcosh}\sqrt{x/y})\, dy}$$

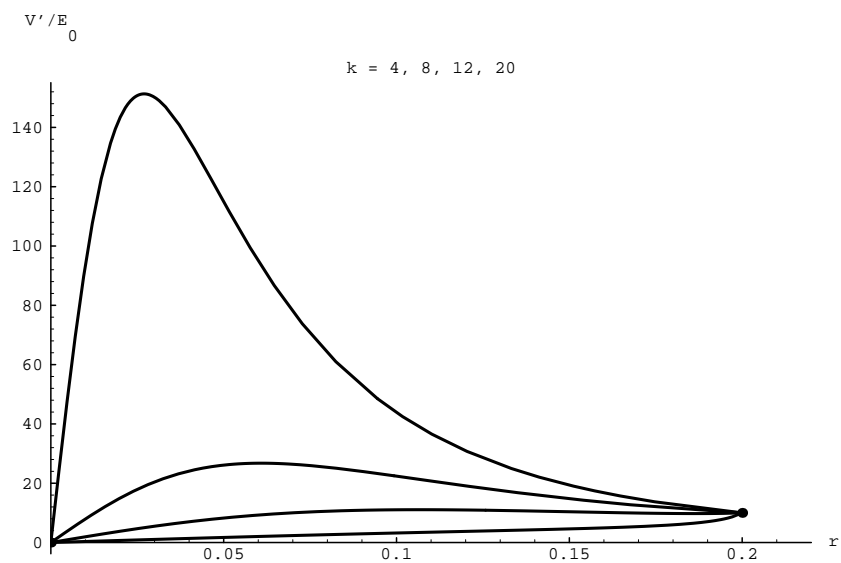

FIGURE 8. First derivative of the potential $V_{Z_K}$ for some values of $k$.

where

$$R_{r^*}(y) = \int_{r^*}^{b_{\max}} \frac{dr}{r^2 \sqrt{yv(r) - r^{-2}}} \tag{60}$$

**B.5. Motion in the potential $V_{Z_K}$.** Integrating the equations of motion for impact parameter $0 \leq b < b_{\max}$ yields the following trajectories in polar coordinates $(r, \phi)$ as functions of the parameter $s \in [b, \infty[$ (there is a singularity at $b = b_{\max}$):

$$\phi(r(s)) = \int_{r_{\min}}^{r(s)} \frac{dz}{z^2 \sqrt{(1 - \frac{V_{Z_K}(z)}{E_0})b^{-2} - z^{-2}}}$$

$$\tag{61} = \begin{cases} -\arccos\left(\frac{b}{s}\right) - \frac{b\,k}{2\pi} \arctan\left(2\frac{\sqrt{(s^2-b^2)(b_{\max}^2-s^2)}}{b^2+b_{\max}^2-2s^2}\right) \\ \qquad\qquad\qquad\qquad \text{if } b \leq s < s_{\text{sing}} \\ \\ -\arccos\left(\frac{\sqrt{2}\,b}{\sqrt{b^2+b_{\max}^2}}\right) - \frac{b\,k}{4} \qquad \text{if } s = s_{\text{sing}} \\ \\ -\arccos\left(\frac{b}{s}\right) - \frac{b\,k}{2\pi} \arctan\left(2\frac{\sqrt{(s^2-b^2)(b_{\max}^2-s^2)}}{b^2+b_{\max}^2-2s^2}\right) - \frac{b\,k}{2} \\ \qquad\qquad\qquad\qquad \text{if } s_{\text{sing}} < s \leq b_{\max} \\ \\ -\arccos\left(\frac{b}{s}\right) - \frac{b\,k}{2} \qquad\qquad \text{if } s > b_{\max} \end{cases}$$

where $s_{\text{sing}} = (b^2 + b_{\max}^2)^{1/2}/\sqrt{2}$ is the singularity in the argument of the arctan.

The trajectories are continuously differentiable everywhere, e.g. for $r = s = b_{\max}$:

$$\lim_{r \to b_{\max}+0} \phi'(r) = \lim_{r \to b_{\max}-0} \phi'(r) = -\frac{b}{b_{\max}\sqrt{b_{\max}^2 - b^2}} \tag{62}$$

Some sample trajectories are shown in figure 4. æ

## Appendix C. Formal Languages and Turing machines

In this section we introduce notions which will serve to characterize the structural complexity of subshifts.

**C.1. Basic definitions.** We start from a finite set $\mathring{A}$ of abstract symbols, which will be called *alphabet* in the following. Define the set of all words or (finite) strings $\mathring{A}^* = \bigcup_{n=0}^{\infty} \mathring{A}^n$. It is called the *total language* over $\mathring{A}$ and together with the concatenation operation becomes a semi-group. Its unit element is the empty string $. Powers correspond to repetitions of symbols, the length $\lg(s)$ of a string $s \in \mathring{A}^*$ yields a formal logarithm. Any subset $L$ of the total language $\mathring{A}^*$ is called a *language*, e.g. $A^+ := A^* \setminus \{\$\}$. A word $x \in L$ is a subword or *segment* of a word $y \in L$, written $x \prec y$ if $x$ appears within $y$, i.e. if there are $a, b \in \mathring{A}^*$ s.t. $y = axb$.

**C.2. Automata.** Automata are mathematical models of devices that process information by giving responses to inputs. Formal language theory views them as scanning devices or *acceptors* able to recognize, whether a word belongs to a given formal language. There is a hierarchy of automata corresponding to the complexity of their structure, in particular of their memory. Let $Z$ be a finite nonempty set of internal states $z_i$, among them an initial state $z_0$, and a nonempty set of final states $Z_f \subset Z$. Let $\mathring{A}$ and $\mathring{W} \supset \mathring{A}$ be two (as usual nonempty, finite) alphabets, called the input alphabet and the working alphabet respectively. The working alphabet contains at least one extra symbol, the blank. We list the following basic types $i = 0, 1, 2, 3$:

- **(0) Turing machine (TM):** A (deterministic) TM given by the tuple $M = (Z, \mathring{A}, \mathring{W}, \omega, \tau, \mu, z_0, Z_f)$ is defined by three transition maps, an *overwriting map* $\omega : Z \times \mathring{W} \to \mathring{W}$, which overwrites the symbol being read, a *next state map* $\tau : Z \times \mathring{W} \to Z$, which defines the machine's transition to a new internal state, and a *head moving map* $\mu : Z \times \mathring{W} \to \{L, R\}$, which moves the reading head of the machine by one position to the right or left. The language accepted by $M$ is the set of words $w \in \mathring{A}^*$ which make $M$, starting from the first letter of $w$, reach a final state.
- **(1) Linear bounded automaton (LBA):** A LBA is a Turing machine whose work space, where its head is allowed to move, is restricted to (or equivalently bounded by a linear function of ) the space the input word uses on the tape.
- **(2) Pushdown automaton (PA):** A PA uses a stack (first in – last out memory) instead of a tape as memory.
- **(3) Finite deterministic automaton (FDA):** A FDA does not have an extra memory. It is given by a single function, namely the next state or transition function $\tau$, i.e. the FDA performs the following actions: It scans the letter $s_i$ of the word $w$, while in an internal state $z \in Z$, it performs a transition to the new state $z' = \tau(z, s_i)$, and then it moves to the next letter $s_{i+1}$. This is repeated until the last letter of $w$ is reached. If the last state is a final state, the word is accepted.

A hierarchy of automata corresponds to the Chomsky hierarchy of languages, namely the recursively enumerable (0), context sensitive (1), context free (2) and regular (3) languages, in the sense that any class i language is recognized by a class i automaton and conversely any class i automaton produces a class i language as output.

TRUNCATED HORSESHOES AND FORMAL LANGUAGES 39Carol., Math. et Phys., to appear in spring 1994, preprint available as chao-dyn/9305008 from nlin-sys@xyz.lanl.gov.
20. P. Walters, *An introduction to ergodic theory*, Springer Verlag, New York, 1982.Technische Universität Berlin, FB Mathematik MA 7-2, Str. d. 17. Juni 136, D-10623 Berlin, Germany
*E-mail address*: troll@math.tu-berlin.de, *FAX:* + 30 - 314 21577